\documentclass[final]{cvpr}

\usepackage{times}
\usepackage{epsfig}
\usepackage{graphicx}
\usepackage{amsmath}
\usepackage{amssymb}
\usepackage{multirow}
\usepackage{multicol}
\usepackage{amsmath}
\usepackage{float}

\usepackage[pagebackref=true,breaklinks=true,colorlinks,bookmarks=false]{hyperref}

\def\cvprPaperID{71} 
\def\confYear{CVPR 2021}

\begin{document}

\title{Efficient Space-time Video Super Resolution using Low-Resolution Flow and Mask Upsampling}

\author{Saikat Dutta\\
IIT Madras, India\\
{\tt\small saikat.dutta779@gmail.com}
\and
Nisarg A. Shah\\
IIT Jodhpur, India\\
{\tt\small shah.2@iitj.ac.in}
\and 
Anurag Mittal\\
IIT Madras, India\\
{\tt\small amittal@cse.iitm.ac.in}
}

\maketitle

\begin{abstract}
This paper explores an efficient solution for Space-time Super-Resolution, aiming to generate High-resolution Slow-motion videos from Low Resolution and Low Frame rate videos. A simplistic solution is the sequential running of Video Super Resolution and Video Frame interpolation models. However, this type of solutions are memory inefficient, have high inference time, and could not make the proper use of space-time relation property. To this extent, we first interpolate in LR space using quadratic modeling. Input LR frames are super-resolved using a state-of-the-art Video Super-Resolution method. Flowmaps and blending mask which are used to synthesize LR interpolated frame is reused in HR space using bilinear upsampling. This leads to a coarse estimate of HR intermediate frame which often contains artifacts along motion boundaries. We use a refinement network to improve the quality of HR intermediate frame via residual learning. Our model is lightweight and performs better than current state-of-the-art models in REDS STSR Validation set.
   
\end{abstract}

\section{Introduction}
With the easy availability of high resolution (HR) displays such as UHD TVs and monitors, the need for visual content to be available at higher resolution is also growing exponentially. However, the video quality in terms of resolution is not available up to the mark of available displays. For instance, most of the visual content available has a resolution of 1080p at 30 FPS or lower, while UHD displays support a resolution of 8K and 120 FPS. Hence, there is enormous scope in the task of translating content to high space-time resolution video from the corresponding lower resolution video. Its application is not only limited to high-definition television but also has it in sports and security applications.
It can also be used as a compression-decompression framework.

Deep neural networks have shown promising results on various video manipulation tasks like Video Super resolution (VSR) \cite{tian2020tdan,isobe2020video}, Video Frame Interpolation (VFI) \cite{kalluri2020flavr,dain}, and Video Deblurring \cite{nah2020ntire} with better computing power availability. 
In Video Super Resolution, we try to increase the spatial resolution of an input video sequence. On the other hand, we aim to increase temporal dimension of an input video in Video Frame Interpolation by inserting new frames between the existing frames.
In Space-time Video Super-Resolution (STSR), our goal is to increase both input video data's spatial and temporal dimension. One of the ways could be sequentially combining VSR and VFI models in a two-stage network. However, time and space are certainly related, and the sequential models could not exploit this property completely, leading to marginal results. Also, predicting high-quality frames requires state-of-art, heavy VSR and VFI models, leading to computationally expensive models.

In this paper, we have presented an efficient framework for Joint Video Super Resolution and Frame Interpolation. Unlike prior work, we have considered non-linear motion between LR frames explicitly through quadratic modeling to interpolate in LR frames. We have used a state-of-the-art Recurrent Neural Network to super-resolve the input LR frames. We have reused intermediate LR flowmaps and blending masks in HR space by using bilinear interpolation rather than directly estimating them in HR space, hence making the method computationally efficient. Estimated HR frames, coarse HR flowmaps and mask produces a coarse intermediate frame estimate. This coarse estimate is further refined by a refinement module. In this work, we have considered 4x upscaling in spatial domain and 2x upscaling in temporal domain. However, our algorithm can be extended to upscaling by any factor in temporal domain.


\section{Related Work}
In this section, we briefly review the literature on related topics i.e. Video Super-Resolution and Video Frame Interpolation, then we proceed to discuss state-of-the-art Spatio-Temporal Video Super-Resolution algorithms.

\subsection{Video Super Resolution}
Video super-resolution is the task of reconstructing a High-resolution video frame from its corresponding Low-resolution frames.
Amidst the success of deep-learning-based methods, specifically in the domain of computer vision, several single-image SR models have been developed.
Some of these methods improve the spatial resolution by concentrating only on the corresponding LR image's spatial information \cite{dong2015image, zhang2019image, lim2017enhanced}. 

However, it is observed that if single-image SR models are applied independently over each frame of the video, then the generated HR video lack temporal consistency, in turn generating flickering effects \cite{shi2016real}.
Therefore several methods are proposed to exploit the temporal relationships for better results and among them, two are very common. First is the simple concatenation of few sequential LR input frames, and second is the use of 3d convolution filters \cite{caballero2017real, jo2018deep, huang2017video, li2019fast}. But, the method of concatenation of frames fail when there large motion displacements or multiple localized motion  \cite{fisr_aaai}. Similarly, 3D convolution increases the computational complexity to an extent which may lead to reduction in accuracy when working in resource constrained environment. 

Some VSR methods use optical flow for temporal alignment. In these methods first they approximate motion by calculating optical flow between the corresponding frame and every neighboring frames. After that they warp the neighboring frames based on predicted motion map \cite{tao2017detail, caballero2017real}.
Muhammad et al. \cite{haris2019recurrent} used iterative refinement framework, which concatenate the input frames with supporting frames multiple times. They computed a residual image for each time step to reduce the error between the expected image and the prediction using the idea of back-projection. Nevertheless, it is not easy to obtain accurate flow and flow warping also introduces artifacts into the aligned frames. This type of effect was solved to an extent by Jo et al. \cite{jo2018deep} using dynamic upsampling.
They mainly used the learned residual image to enhance the sharpness and took advantage of the network's captured implicit motion. 

Additionally, Tian et al. \cite{tian2020tdan} proposed TDAN for temporal alignment without estimation of motion using deformable alignment. Wang et al. \cite{wang2019edvr} proposed EDVR which further explores usage of multi-scale information in TDAN. Later, aligned frames are fused using temporal and spatial attention mechanism.
While, in RSDN \cite{isobe2020video}, the input is divided into structure and detail components and later fed into recurrent unit made up of structural-detail blocks. This method is lightweight and effective in exploiting information from prior frames for super-resolution; therefore, we have used this network for VSR in our proposed framework. 
\begin{figure*}[htbp]
    \centering
    \includegraphics[width=0.9\textwidth]{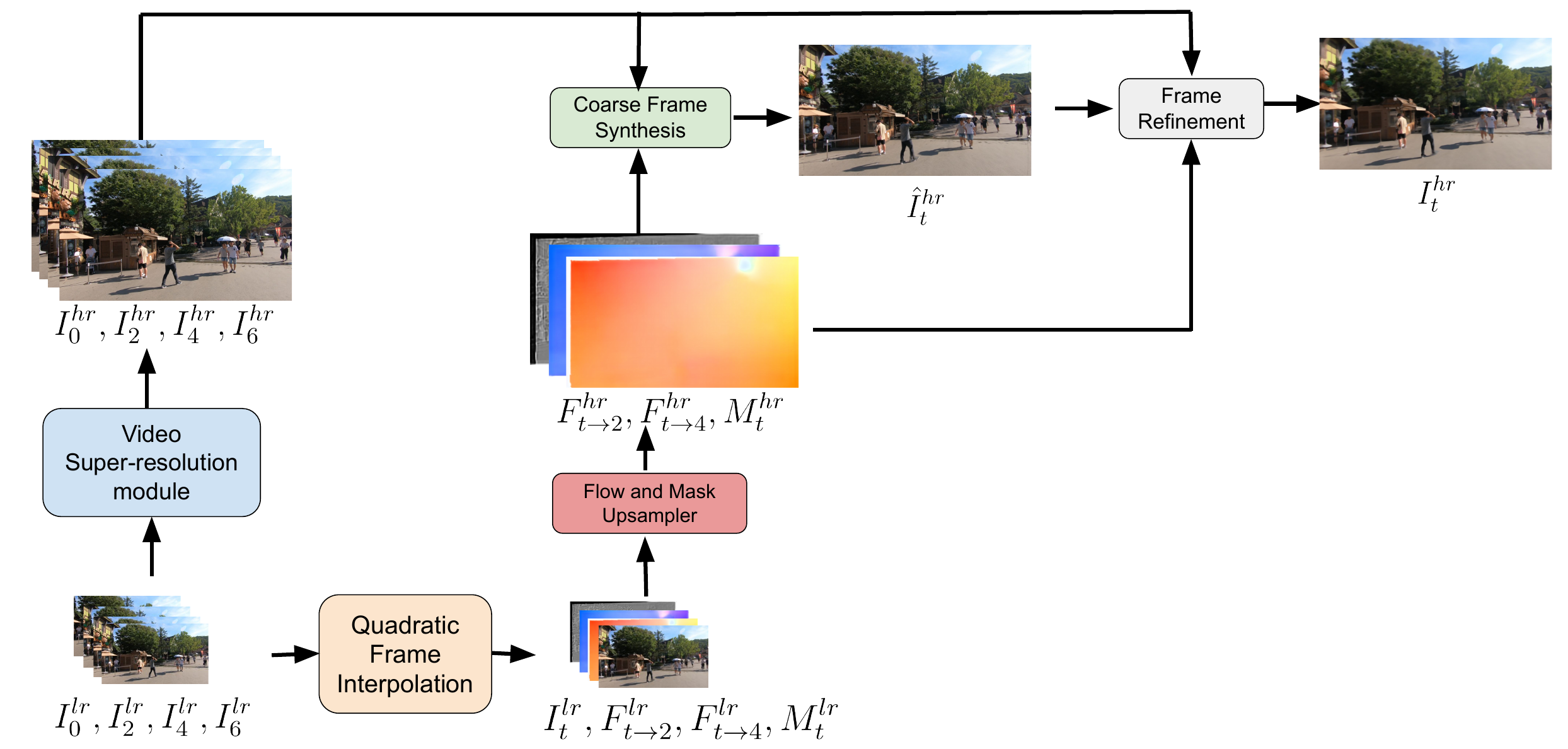}
    \caption{Overview of our Space-Time Super Resolution Framework.}
    \label{main_diag}
\end{figure*}
\begin{figure*}
    \centering
    \includegraphics[width=0.9\textwidth]{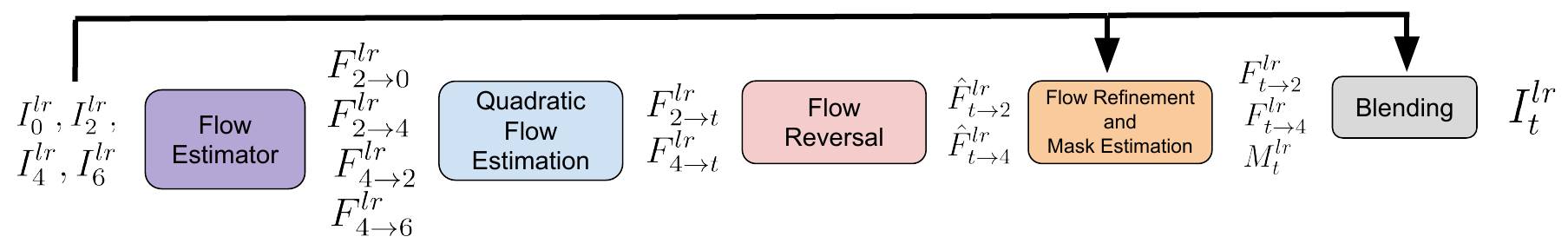}
    \caption{LR Frame Interpolation Framework: Quadratic Frame Interpolation.}
    \label{qvi_diag}
\end{figure*}
\subsection{Video Frame Interpolation}
Early works in Video Frame Interpolation are often based on Optical Flow estimation, and interpolation accuracy is used to compute the quality of optical flow \cite{baker2011database,barron1994performance}. Herbst et al.\cite{herbst2009occlusion} use bidirectional flow to estimate intermediate flow and perform occlusion reasoning. Using intermediate flows and occlusion masks, they generate the final frame by a blending algorithm. 

Long et al. \cite{mind} uses a Deep encoder-decoder architecture to directly synthesize interpolated frame from two consecutive video frames. Liu et al. \cite{dvf} computes voxel flow from two input frames by a fully convolutional network and interpolated frame is generated by trilinear interpolation.
Super-SloMo\cite{superslomo} estimates bi-directional optical flows using a U-Net \cite{unet} and predicts intermediate optical flows and soft visibility maps using another U-Net. Finally, they fuse warped frames linearly to generate the intermediate frame. Niklaus et al. \cite{ctxsyn} estimate per-pixel context maps from pre-trained ResNet and use warped context maps for frame synthesis. Liu et al. \cite{cyclicgen} introduce Cycle-consistency loss and use edge maps to improve over Deep Voxel Flow \cite{dvf}. Xue et al. \cite{toflow} learned self-supervised task-specific optical flow for various Video enhancement problems, including temporal interpolation.

Niklaus et al. \cite{adaconv} learns spatially adaptive kernels for each pixel in the interpolated frame using a fully convolutional neural network. The same authors estimated two separable 1D kernels per each pixel in \cite{sepconv} to reduce computational complexity and improve performance. MEMC-Net \cite{memc} uses Adaptive Warping Layer that utilizes both optical flow and interpolation kernels to synthesize the target frame. Bao et al. \cite{dain} propose Depth-Aware Flow Projection layer to estimate intermediate optical flows. 

Recently, researchers leverage more than two input frames to capture non-linear motion between frames. Xu et al. \cite{qvi} assume quadratic motion of the pixels and show improvement upon linear models. Kalluri et al. \cite{kalluri2020flavr} use a 3D UNet architecture to generate interpolated frames from four input RGB frames without help of any extra information like optical flow or depth.

\subsection{Spatio-temporal Video Super Resolution} Kim et al. \cite{fisr_aaai} propose a joint VFI-SR framework to increase both spatial and temporal dimension by a factor of 2. 
The framework incorporates novel temporal loss at multiple scale working as temporal regularizer on the input sequence.
STARnet framework proposed by Haris et al. \cite{starnet}
consists of three stages. In first stage, both LR and HR feature maps are learnt for existing and intermediate frames along with a motion representation from input LR frames and bidirectional optical flowmaps. The HR and LR feature maps are further refined in Stage-2, while Stage-3 reconstructs corresponding HR and LR frames from the feature maps. Kang et al. \cite{stvun_eccv} uses an encoder to get feature representation of each input frame. These encoded features are fused using ``Early Fusion with Spatio-Temporal weights" (EFST) module for Spatial upsampling. The encoded features are interpolated using computed optical flow for temporal upsampling. Finally, decoder block computes residues for both spatial and temporal upsampling. Xiang et al. \cite{zsm_cvpr} extracts feature from each input LR frames and feeds the extracted feature maps to Frame Feature interpolation module to synthesize intermediate feature maps in LR space. Now consecutive LR feature maps are passed to a Bidirectional Deformable ConvLSTM module for temporal context aggregation. Finally, the output feature maps from ConvLSTM module are passed to a Frame Reconstruction module to generate the final output frames.

\begin{figure*}
    \centering
    \includegraphics[width=0.9\textwidth]{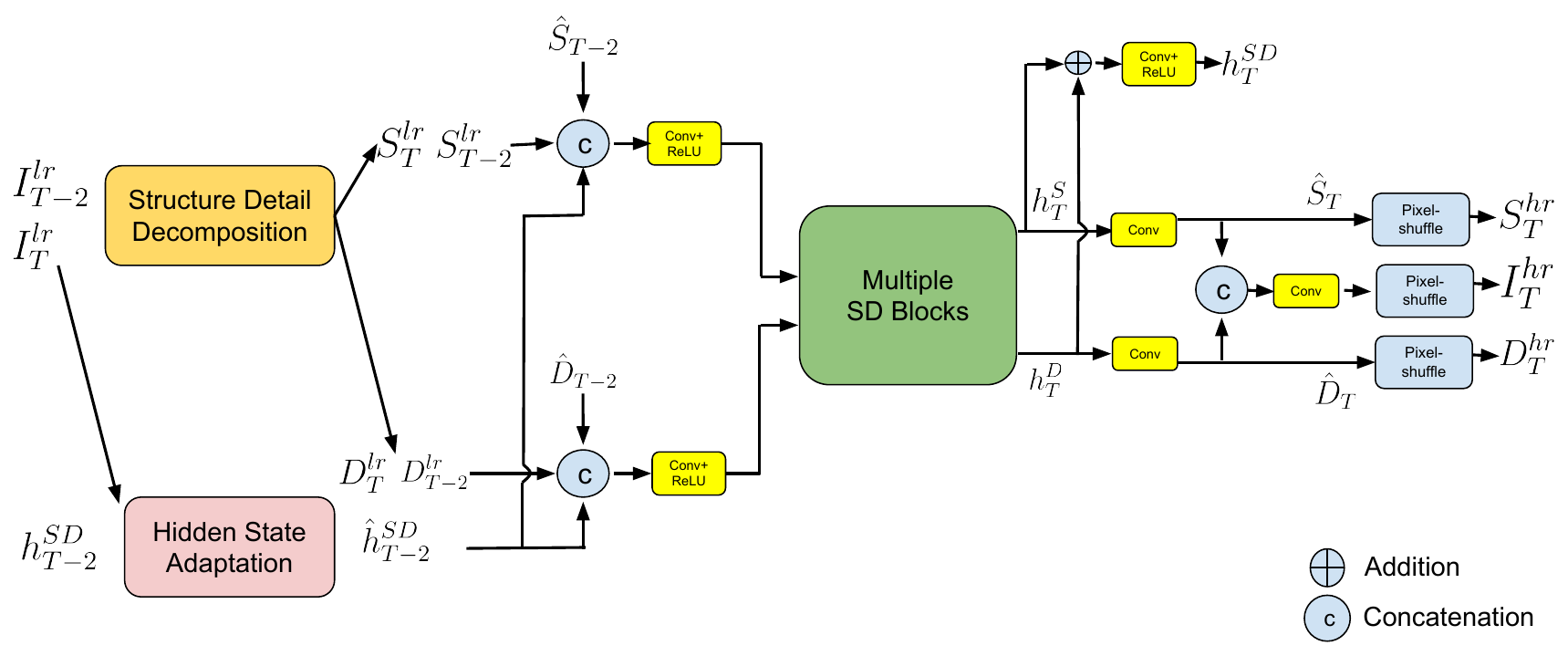}
    \caption{VSR Framework: Recurrent Structure-Detail Network.}
    \label{rsdn_diag}
\end{figure*}
\section{Proposed Method}
Given input Low Resolution-Low Frame Rate (LR-LFR) frames $I^{lr}_{0}$, $I^{lr}_{2}$, $I^{lr}_{4}$ and $I^{lr}_{6}$, our goal is to predict High Resolution-High Frame Rate (HR-HFR) frames $I^{hr}_{2}$, $I^{hr}_{t}$ and $I^{hr}_{4}$, where $t \in (2,4)$. In this work, we have aimed to synthesize only one intermediate frame i.e. $t=3$, however our method can be extended to produce multiple intermediate frames. Our model pipeline consists of three parts: (a) LR Frame Interpolation, (b) HR Frame Reconstruction and (c) HR Intermediate Frame Reconstruction. We describe each of these stages in detail in the following. The model diagram is shown in Fig. \ref{main_diag}.

\subsection{LR Frame Interpolation:}
For Frame interpolation in LR space, we use Quadratic Frame Interpolation (QFI) \cite{qvi}. Unlike many state-of-the-art Video Frame Interpolation methods \cite{dvf,ctxsyn,superslomo,dain}, QFI uses four frames to model non-linear motion. First, a flow estimation module is used to compute flowmaps between neighbor frames ($F^{lr}_{2\rightarrow0},F^{lr}_{2\rightarrow4},F^{lr}_{4\rightarrow2},F^{lr}_{4\rightarrow6}$). PWCNet \cite{pwcnet} is used as flow estimator in this work. Assuming quadratic motion between frames, intermediate flow maps $F^{lr}_{2\rightarrow t}$ and $F^{lr}_{4\rightarrow t}$ are given by\footnote{We refer the reader to \cite{qvi} for derivation.},
\begin{equation}
\small
\begin{split}
    F^{lr}_{2\rightarrow t} = 0.5 \times (F^{lr}_{2\rightarrow4}+ F^{lr}_{2\rightarrow0})\times (\tfrac{t-2}{2})^2 + \\ 
     0.5 \times (F^{lr}_{2\rightarrow4}- F^{lr}_{2\rightarrow0})\times (\tfrac{t-2}{2})
\end{split}
\end{equation}
\begin{equation}
\small
\begin{split}
    F^{lr}_{4\rightarrow t} = 0.5 \times (F^{lr}_{4\rightarrow2}+ F^{lr}_{4\rightarrow6})\times (\tfrac{4-t}{2})^2 + \\ 
     0.5 \times (F^{lr}_{4\rightarrow2}- F^{lr}_{4\rightarrow6})\times (\tfrac{4-t}{2})
\end{split}
\end{equation}

These intermediate flowmaps are passed to a flow reversal layer to generate $F^{lr}_{t\rightarrow 2}$ and $F^{lr}_{t\rightarrow 4}$. These estimated flowmaps often contain ringing artifacts which are refined by the help of a flow refinement module. Unlike in QFI, we use Gridnet \cite{gridnet,ctxsyn} as Flow Refinement module. Further, we use a 3-layer network to generate blending mask $M_t^{lr}$. The blending mask $M_t^{lr}$ helps us in blending warped frames to generate intermediate frame. Finally, the LR intermediate frame $I^{lr}_{t}$ is synthesized as, 

\begin{equation}
\small
\begin{split}
    I^{lr}_{t} = \frac{ \tfrac{4-t}{2} \times M_t^{lr} \odot bw(I^{lr}_2, F^{lr}_{t\rightarrow2}) } 
    {\tfrac{4-t}{2} \times M_t^{lr}+\tfrac{t-2}{2} \times (1-M_t^{lr})}
    + \\
    \frac{\tfrac{t-2}{2} \times (1-M_t^{lr}) \odot bw(I^{lr}_4, F^{lr}_{t\rightarrow4})}
    {\tfrac{4-t}{2} \times M_t^{lr} + \tfrac{t-2}{2} \times (1-M_t^{lr})}
    \label{lr_syn}
\end{split}
\end{equation}
where $bw(.,.)$ is the backward warping function and $\odot$ denotes hadamard product. The overall diagram of LR frame interpolation is shown in Fig. \ref{qvi_diag}.

\subsection{HR Frame Reconstruction:}
We use a state-of-the-art Video Super Resolution method, Recurrent Structure Detail Network (RSDN) \cite{isobe2020video} for generating \{$I^{hr}_0$, $I^{hr}_2$, $I^{hr}_4$, $I^{hr}_6$\} from corresponding LR frames \{$I^{lr}_0$, $I^{lr}_2$, $I^{lr}_4$, $I^{lr}_6$\}. Please note that, we do not use LR interpolated frame $I^{lr}_t$ as input to RSDN, since the inaccuracy in interpolation can affect super-resolution performance. For the sake of completeness, we discuss RSDN in the following. Model diagram of RSDN is shown in Fig. \ref{rsdn_diag}.

RSDN is a recurrent neural network, which works on Structure and Detail components on input frames rather than the whole frames. Structure component and Detail components capture low-frequency and high-frequency information in the images respectively. Structure and Detail components are processed by two similar parallel branches. The Detail branch is explained below.

At a given time step $T$, Hidden State Adaptation (HSA) module adapts previous hidden state $h^{SD}_{T-2}$ according to current frame $I^{lr}_T$ to produce adapted hidden state $\hat{h}^{SD}_{T-2}$. Detail components of previous and current frames \{$D^{lr}_{T-2}$, $D^{lr}_{T}$\} are concatenated along with estimated detail map of previous frame $\hat{D}_{T-2}$ and adapted hidden state $\hat{h}^{SD}_{T-2}$. The concatenated feature maps are further passed to a convolutional layer and a number of Structure-Detail (SD) blocks to produce $h^{D}_{T}$. SD blocks are modified residual blocks which can fuse information from structure and detail branches effectively. $h^{D}_{T}$ is fed to a convolutional layer and an upsampling layer to produce Detail map at current time step, $\hat{D}_{T}$. Similarly, $h^{S}_{T}$ and $\hat{S}_{T}$ is generated in the Structure branch. $h^{S}_{T}$ and $h^{D}_{T}$ are combined by convolutional layers to generate $I^{hr}_T$ and hidden state at current time step, $h^{SD}_T$.
\subsection{HR Intermediate Frame Reconstruction:}
Solving VFI in HR space is computationally expensive mostly because of Flow estimation module inside our VFI framework. PWCNet takes $0.026$ seconds to compute flow between a pair of LR ($180\times320$) frames, whereas it takes 3.4x runtime ($0.089$ seconds) to process a pair of HR ($720\times1280$) frames. We should also note that QFI computes flowmaps between four pairs of frames, hence the overall runtime of the model increases by a large margin.
In addition to that, in HR space motion will be large, so flow estimation module will struggle to find pixel correspondences accurately, therefore creating a performance bottleneck. Instead, we reuse intermediate flowmaps and mask from LR space in this work. 
We upscale LR intermediate flowmap and blending mask with the help of bilinear interpolation.
The coarse HR intermediate flowmap and mask estimates are given by, 

\begin{equation}
    F^{hr}_{t\rightarrow2} = 4 \times up(F^{lr}_{t\rightarrow2})
\end{equation} 
\begin{equation}
    F^{hr}_{t\rightarrow4} = 4 \times up(F^{lr}_{t\rightarrow4})
\end{equation} 
\begin{equation}    
    M_t^{hr} = up(M_t^{lr})
\end{equation}
where $up(.)$ denotes bilinear upsampling by a factor of 4.

Similar to Equation-\ref{lr_syn}, we can produce a coarse estimate for intermediate HR frame using predicted HR frames and upscaled flow and masks. Hence, the coarse estimate for intermediate frame $\hat{I}^{hr}$ is given by,

\begin{equation}
\small
\begin{split}
 \hat{I}^{hr}_{t} = \frac{ \tfrac{4-t}{2} \times M_t^{hr} \odot bw(I^{hr}_2, F^{hr}_{t\rightarrow2})} 
    {\tfrac{4-t}{2} \times M_t^{hr} + \tfrac{t-2}{2} \times (1-M_t^{hr})}
    + \\
    \frac{\tfrac{t-2}{2} \times (1-M_t^{hr}) \odot bw(I^{lr}_4, F^{hr}_{t\rightarrow4})}  
    {\tfrac{4-t}{2} \times M_t^{hr} + \tfrac{t-2}{2} \times (1-M_t^{hr})}
\label{hr_syn}
\end{split}
\end{equation}

Since the coarse estimate obtained by Equation-\ref{hr_syn} depends on upscaled flow and masks, we can further refine this coarse estimate with the help of a refinement module. We have used Gridnet as the refinement network in this work. Estimated HR frames, upscaled flowmaps and blending mask, warped HR frames, coarse estimate of intermediate HR frame is fed to the refinement network. The refinement network learns a residual image with respect to the coarse estimate of intermediate HR frame. The final estimate of intermediate HR frame is given by, 
\begin{equation}
\begin{split}
    I^{hr} = \hat{I}^{hr}_t + ref(\hat{I}^{hr}_t, I^{hr}_2,  I^{hr}_4, F^{hr}_{t\rightarrow2}, 
    F^{hr}_{t\rightarrow4}, M^{hr}_t, \\
    bw(I^{hr}_2,F^{hr}_{t\rightarrow2}), bw(I^{hr}_4,F^{hr}_{t\rightarrow4}) )
\end{split}    
\end{equation}
where ``$ref$" is the Frame Refinement module.

\begin{figure*}[h]
    \centering
    \includegraphics[width=0.9\textwidth]{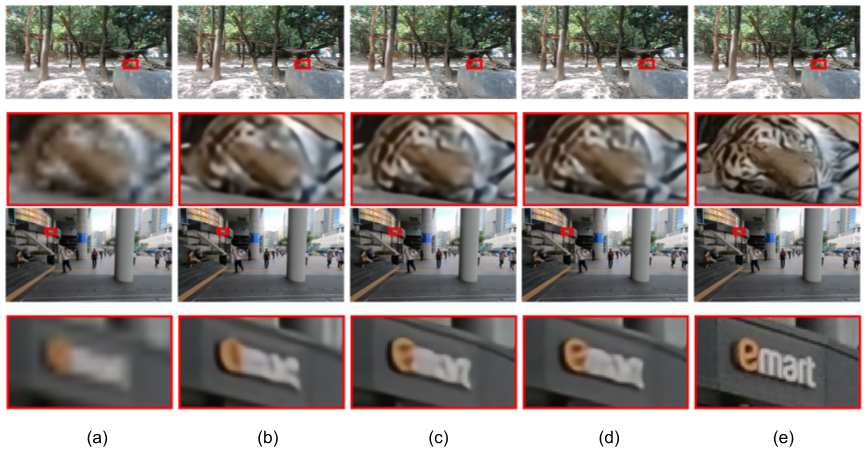}
    \caption{Comparison with state-of-the-art on Even Frame generation. From Left: (a) Upsampled LR frames (b) STARnet (c) Zooming Slomo (d) Ours (e) Ground Truth. Zoom in for details.}
    \label{sota_even}
\end{figure*}

\begin{figure*}[h]
    \centering
    \includegraphics[width=0.9\textwidth]{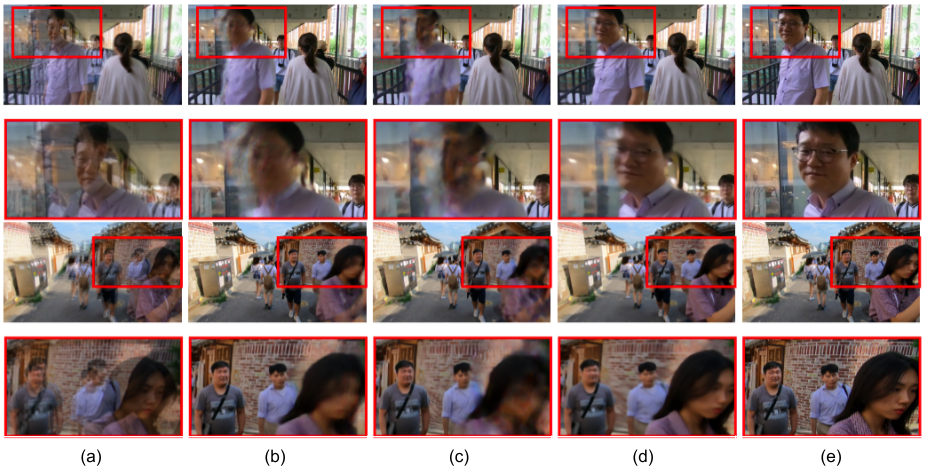}
    \caption{Comparison with state-of-the-art on Odd Frame generation. From Left: (a) Upsampled and overlayed LR frames (b) STARnet (c) Zooming Slomo (d) Ours (e) Ground Truth. Zoom in for details.}
    \label{sota_odd}
\end{figure*}

\section{Experiments}
\subsection{Dataset Description}
 We have used REDS STSR dataset \cite{STSR_challenge} for training our models. REDS STSR dataset contains a variety of dynamic scenes. The training split contains 30 video sequences, where each sequence contains 100 frames each. Validation and Test splits contains 30 sequences each. HR image resolution of this dataset is $720 \times 1280$. LR frames are generated by 4x Bicubic downsampling. Since the HR-HFR frames of test split are not publicly available yet, we use the Validation split for evaluation purpose. We use even LR frames as input and predict all the HR-HFR frames during evaluation (starting index is 0).

\subsection{Training Details}
We have implemented our models in Python with Pytorch \cite{pytorch} framework on a system with one NVIDIA 1080 Ti GPU. We have used Adam optimizer with $\beta_1=0.9$ and  $\beta_2=0.99$ with a batch size of 2. Initial learning rate is set to $2\times10^{-4}$ and gradually reduced to $2\times10^{-6}$. Patches of size $128 \times 128$ are cropped randomly during training. Frame sequences are randomly flipped horizontally and vertically along with random temporal order reversal during training. 

We have used pretrained PWCNet and RSDN and only finetune these modules with a low learning rate of $2\times10^{-6}$ at a later stage in training.

\subsection{Loss Functions}

\textbf{Frame Reconstruction loss:} We have used Charbonnier loss between predicted frames and ground truth frames. Frame reconstruction loss is utilized for LR intermediate frame alongside HR frames, since accurate reconstruction of LR intermediate frame can help the network to reconstruct HR intermediate frame well. Frame Reconstruction loss $\mathcal{L}_{fr}$ is given by,

\begin{equation}
\small
\mathcal{L}_{fr} = \sum_{i \in \{0,2,t,4,6\} }\mathcal{L}_c(I^{hr}_i,I^{hr,gt}_i) +   0.5 \times \mathcal{L}_c(I^{lr}_t,I^{lr,gt}_t)
\end{equation}
where $\mathcal{L}_c(x,y) = \sqrt{||x-y||^2 + \epsilon^2}$ denotes Charbonnier loss \cite{charbonnier}. We have used $\epsilon=0.001$ in our experiments.

\textbf{Structure-Detail loss:} Inspired by \cite{isobe2020video}, we want to put emphasis on both structure and detail components of the reconstructed HR frames. We have used Charbonnier loss on both structure and detail components for this purpose. Structure-Detail loss is given by,

\begin{equation}
\small
    \mathcal{L}_{sd} = \sum_{i \in \{0,2,t,4,6\} }\mathcal{L}_c(S^{hr}_i,S^{hr,gt}_i) +  \sum_{i \in \{0,2,t,4,6\} }\mathcal{L}_c(D^{hr}_i,D^{hr,gt}_i)
\end{equation}
where ``$S$" and ``$D$" denotes corresponding structure and detail components respectively.

Our final loss function is given by,
\begin{equation}
    \mathcal{L} = \alpha_{fr} \mathcal{L}_{fr} + \alpha_{sd} \mathcal{L}_{sd}
\end{equation}
We have used $\alpha_{fr}$ = $\alpha_{sd}$ = 45 in our experiments.

\begin{figure*}[h]
    \centering
    \includegraphics[width=0.85\textwidth]{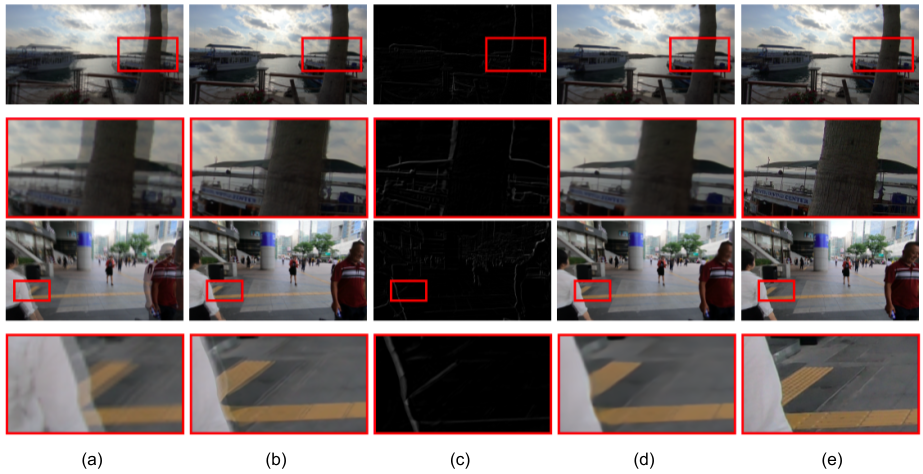}
    \caption{Importance of HR Frame Refinement. From left: (a) Upsampled and overlayed LR frames (b) Ours (w/o refinement) (c) Residual map calculated by Frame Refinement module (d) Ours (w/ refinement) (e) Ground Truth. Zoom in for details.}
    \label{res_abl}
\end{figure*}

\subsection{Results}
\subsubsection{Evaluation Metrics} 
We have used Peak Signal-to-Noise Ratio (PSNR) and Structural Similarity Index (SSIM) \cite{ssim} as evaluation metrics in our experiments. 
\begin{table*}[!h]
\begin{tabular}{|c|c|c|c|c|c|c|}
\hline
\multirow{2}{*}{\textbf{Method}} & \multicolumn{2}{c|}{\textbf{Even Frames}} & \multicolumn{2}{c|}{\textbf{Odd Frames}} & \multicolumn{2}{c|}{\textbf{Overall}} \\ \cline{2-7} 
                                 & \textbf{PSNR}       & \textbf{SSIM}       & \textbf{PSNR}       & \textbf{SSIM}      & \textbf{PSNR}     & \textbf{SSIM}     \\ \hline
STARnet \cite{starnet}                  &      28.43               &        0.7978             &        21.55             &      0.5925              &       25.03            &       0.6961            \\ \hline
Zooming Slomo \cite{zsm_cvpr}            &      \textbf{28.95}               &       \textbf{0.8151}              &     21.63                &          0.6010          &             25.33      &       0.7091            \\ \hline
Ours                             &           28.56          &     0.8018                &         \textbf{22.41}            &         \textbf{0.6148}           &     \textbf{25.51}              &         \textbf{0.7093}          \\ \hline
\end{tabular}
\centering
\caption{Quantitative comparison with other state-of-the-art models.}
\label{sota_comp}
\end{table*}

\subsubsection{Comparison with state-of-the-art methods}
We have compared our method against two state-of-the-art methods STARnet \cite{starnet} and Zooming Slomo \cite{zsm_cvpr}. We did not compare with STVUN \cite{stvun_eccv}, since the authors used a different degradation model for downsampling frames, hence comparison would not have been fair. We have not compared with FISR \cite{fisr_aaai}, since FISR does spatial upsampling by 2x instead of 4x. We have used pretrained models provided by authors for comparison. We have measured performance on Even frames (VSR) and Odd Frames (VSR+VFI) separately. Quantitative comparison is shown in Table \ref{sota_comp}. Our model achieves significant improvement on PSNR and SSIM scores for Odd frames compared to other two state-of-the-art methods. 
We achieve improvement of 0.78 dB and 0.0138 in PSNR and SSIM respectively compared to the second best algorithm, Zooming Slomo for Odd frames. 
Our overall PSNR and SSIM scores are also better than other algorithms. Qualitative comparison for Even and Odd frames are shown in Fig. \ref{sota_even} and \ref{sota_odd} respectively. Our algorithm produces better results than STARnet in case of Even frames (refer Fig. \ref{sota_even}). From Fig. \ref{sota_odd}, it is clear that our method can handle large motion between frames quite well and performs significantly better than STARnet and Zooming Slomo in Odd frame generation.

\subsubsection{NTIRE 2021 Video Super-Resolution Challenge: Track 2}
We have participated in NTIRE 2021 Video Super-Resolution Challenge: Track 2 (Spatio-temporal) \cite{STSR_challenge}. A total of 223 participants registered in this competition out of which 26 teams participated in the validation phase and 14 teams entered the test phase. The challenge organizers considered PSNR and SSIM scores on REDS STSR test data jointly as primary criteria to release rankings. Our team ranked 10th among the teams participating in the final phase.

\begin{figure*}[h]
    \centering
    \includegraphics[width=\textwidth]{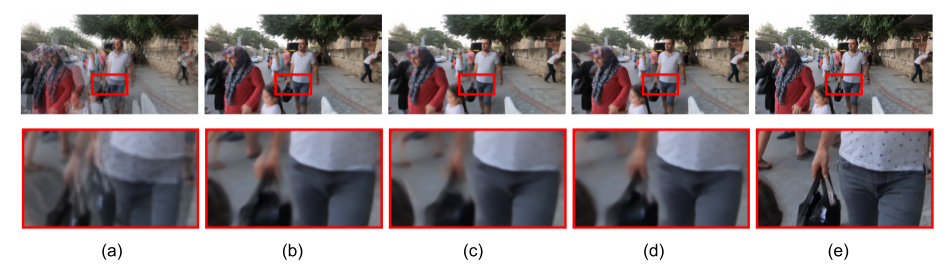}
    \caption{Effect of different architectures in refinement modules. From left: (a) Upsampled and overlayed LR Frames (b) UNet (c) UNet++ (d) Gridnet (e) Ground Truth. Zoom in for details.}
    \label{abl_arch}
\end{figure*}

\subsection{Efficiency}
Our model has 20.09 M parameters. We have reported model sizes of other state-of-the-art methods in Table \ref{runtime}. 
We can see that total number of parameters in our model is $18\%$ of the same in STARnet \cite{starnet} and our model is more lightweight than STVUN \cite{stvun_eccv}.
We have compared our model runtime against state-of-the-art STSR methods. Average runtime required to generate one HR-HFR frame of resolution $720 \times 1280$ in our system is reported in Table \ref{runtime}. We can see that our model is significantly faster than STARnet  and has similar runtime when compared to STVUN. Our model consumes less GPU memory during inference than state-of-the-art Zooming Slomo. 

\begin{table}[h]
\footnotesize
\begin{tabular}{|c|c|c|c|}
\hline
\textbf{\textbf{Method}}      & \textbf{\begin{tabular}[c]{@{}c@{}}Runtime \\ (s)\end{tabular}} & \textbf{\begin{tabular}[c]{@{}c@{}}Parameters \\ (M)\end{tabular}} & \textbf{\begin{tabular}[c]{@{}c@{}}GPU Memory\\ Usage (GB)\end{tabular}} \\ \hline
STARnet \cite{starnet}        & 1.13                                                            & 111.61                                                             & 5.27                                                                     \\ \hline
STVUN \cite{stvun_eccv}       & 0.24                                                            & 30.90                                                              & \textbf{2.90}                                                            \\ \hline
Zooming Slomo \cite{zsm_cvpr} & \textbf{0.15 }                                                  & \textbf{11.10}                                                     & 4.55                                                                     \\ \hline
Ours                          & 0.25                                                            & 20.09                                                              & 3.43                                                                     \\ \hline
\end{tabular}
\caption{Runtime, Parameter and Memory usage consumption comparison with State-of-the-art methods.}
\label{runtime}
\end{table}

\begin{table}[b]
\centering
\small
\begin{tabular}{|c|c|c|}
\hline
\textbf{Method}       & \textbf{PSNR}  & \textbf{SSIM}   \\ \hline
Ours (w/o refinement) & 21.02          & 0.5580          \\ \hline
Ours (w/ refinement)  & \textbf{22.41} & \textbf{0.6148} \\ \hline
\end{tabular}
\caption{Importance of Frame Refinement: Performance comparison on Odd Frames.}
\label{tab:imp_ref}
\end{table}

\begin{table*}[h]
\centering
\small
\begin{tabular}{|c|c|c|c|c|c|c|c|c|}
\hline
\multirow{2}{*}{\textbf{\begin{tabular}[c]{@{}c@{}}Architecture used in \\ Refinement modules\end{tabular}}} & \multicolumn{2}{c|}{\textbf{Even Frames}} & \multicolumn{2}{c|}{\textbf{Odd Frames}} & \multicolumn{2}{c|}{\textbf{Overall}} & \multirow{2}{*}{\textbf{Runtime (s)}} & \multirow{2}{*}{\textbf{\begin{tabular}[c]{@{}c@{}}Total no. of\\ Parameters (M)\end{tabular}}} \\ \cline{2-7}
                                                                                                             & \textbf{PSNR}       & \textbf{SSIM}       & \textbf{PSNR}      & \textbf{SSIM}       & \textbf{PSNR}     & \textbf{SSIM}     &                                       &                                                                                                 \\ \hline
UNet                                                                                                         & \textbf{28.56}      & 0.8015              & 22.38              & 0.6108              & 25.50             & 0.7071            & 0.26                                  & 55.22                                                                                           \\ \hline
UNet++                                                                                                       & \textbf{28.56}      & \textbf{0.8018}     & 22.36              & 0.6121              & 25.49             & 0.7079            & \textbf{0.24}                         & \textbf{17.94}                                                                                  \\ \hline
Gridnet                                                                                                      & \textbf{28.56}      & \textbf{0.8018}     & \textbf{22.41}     & \textbf{0.6148}     & \textbf{25.51}    & \textbf{0.7093}   & 0.25                                  & 20.09                                                                                           \\ \hline
\end{tabular}
\caption{Quantitative comparison between different architectures in refinement modules.}
\label{tab_abl_arch}
\end{table*}

\subsection{Ablation Study}
\subsubsection{Importance of HR Frame Refinement}
Our coarse estimate of HR intermediate frame, $\hat{I}^{hr}_t$ is generated from coarse (upscaled) flow maps and blending masks. Due to 4x upscaling using bilinear interpolation, it is expected that $
F^{hr}_{t\rightarrow2},
F^{hr}_{t\rightarrow4}$ and $M_t^{hr}$ will have inaccuracies along motion boundaries, producing ghosting artifacts in $\hat{I}^{hr}_t$. To address this issue, we have used a refinement module that aims to produce a better estimate of HR intermediate frame through residual learning. To analyze the importance of HR Frame Refinement, we compare the outputs $\hat{I}^{hr}_t$ and $I^{hr}_t$. We denote $\hat{I}^{hr}_t$ as ``Ours (w/o refinement)" and $I^{hr}_t$ as ``Ours (w/ refinement)". Since even frame outputs are independent of this change, we have compared evaluation metrics on odd frames in Table-\ref{tab:imp_ref}. We can infer that we achieve significant improvement on both metrics in generating HR intermediate frames. From Fig. \ref{res_abl}, we can observe that our Frame refinement module performs quite well in removing artifacts from the coarse estimate of intermediate HR frame.

\subsubsection{Choice of architecture in refinement modules}
In addition to Gridnet, we have used UNet \cite{unet} and UNet++ \cite{unet++} in Flow refinement and Frame refinement modules. Details of these architectures can be found in supplementary material.
Our model with Gridnet has $63.6\%$ less parameters than our model with UNet. Our model with UNet++ has $10.7\%$ less parameters than our model with Gridnet. All three models have similar runtimes and our model with Gridnet produces best quantitative results as shown in Table \ref{tab_abl_arch}. Qualitative comparison in Fig \ref{abl_arch} shows our model with Gridnet performs better than other models in generating odd frames.

\section{Conclusion}
In this work, we propose an efficient approach for Space-time Super Resolution. We have adopted a state-of-the-art VSR method RSDN to super-resolve input LR frames. We have used quadratic motion modelling to interpolate in LR space. Flow maps and blending mask from LR space is used to generate a coarse HR intermediate frame estimate. This estimate is further refined by a Frame Refinement network via residual learning. Our model has outperformed existing state-of-the-art models on REDS STSR Validation dataset. We have gained significant improvement on generating HR intermediate frames over other state-of-the-art methods. Our model contains only 20 M parameters and can generate HR-HFR frames in 0.25 seconds on average. We have focused on refining the coarse estimate of HR intermediate frame in this work, however directly refining HR flowmaps and blending masks to generate HR intermediate frame can be tried out as future research direction. Additionally, LR flow maps can be exploited to warp neighboring frames, which can be used as input to VSR module.


{\small
\bibliographystyle{ieee_fullname}
\bibliography{cvpr}
}

\end{document}